# Determination of hybrid and direct laser acceleration dominated regimes in a 55fs laser driven plasma accelerator with ionization induced injection

D. Hazra[1,*], A. Moorti[1,2,†], A. Upadhyay[2], and J. A. Chakera[1,2]

[1]Homi Bhabha National Institute, Training School Complex, Anushakti Nagar, Mumbai-400094, India

[2]Laser Plasma Division, Raja Ramanna Centre for Advanced Technology, Indore-452013, India

[*]E-mail:dhazraphys@gmail.com

†Corresponding Author: moorti@rrcat.gov.in

**Abstract:**

An experimental study on ~55fs (intensity ~$5\times10^{18}$ W/cm$^2$) laser driven plasma accelerator using mixed gas-jet target (He+few%N$_2$) with varying plasma density (~2-7.1×10$^{19}$cm$^{-3}$) is used to identify applicable acceleration mechanisms, viz. hybrid: Direct Laser Acceleration (DLA) + Wakefield, and DLA. Towards lower density of ~$2\times10^{19}$cm$^{-3}$, electron acceleration could be attributed mainly to DLA with ionization induced injection. With increase in density, increasing role of wakefield was observed leading to hybrid regime, and at densities higher than self-injection threshold (≥$5.8\times10^{19}$cm$^{-3}$, observed experimentally for He target) contribution of DLA and wakefield was found to be comparable. Dominant DLA mechanism was also observed in case of pure N$_2$ target with ionization induced injection at a density of ~$2\times10^{19}$cm$^{-3}$. 2D PIC simulations performed using the EPOCH code corroborate the above scenario, and also showed generation of surface waves, considered as a potential mechanism of pre-acceleration to DLA.

**1. Introduction:**

Several electron acceleration mechanisms e.g. Laser wakefield acceleration (LWFA), and Direct Laser Acceleration (DLA) etc. are applicable in high intensity laser plasma interaction. Through LWFA [1], subsequent to generation of high quality quasi-monoenergetic (QM) electron beams [2-4], acceleration of electrons to GeV class energies has been demonstrated [5-11] in the bubble or blowout regime [12,13], where an intense short laser pulse such that $L<\lambda_p$, (where L is the laser pulse length and $\lambda_p$ is the plasma wavelength) is used satisfying a matching condition between laser focal spot ($\omega_0$), normalized laser intensity ($a_0$) and bubble radius (R) such that $k_pR \sim k_p\omega_0=2\sqrt{a_0}$, where $k_p=2\pi/\lambda_p$. Another wakefield mechanism termed as self-modulated laser wakefield acceleration (SMLWFA) is applicable for cases $L>\lambda_p$, achieved at comparatively higher plasma density and longer laser pulse duration. This regime has been studied using several hundreds of fs [14-16] as well as few tens of fs long laser pulses [17-21], and generation of electron beams with broad spectrum and QM spectrum were reported respectively. Generation of QM beams was explained considering long laser pulses are modulated at the order of $\lambda_p$ leading to formation of multiple short laser pulse lets which are intense enough to create bubble regime conditions [17].

In the regime of $L>>\lambda_p$, another possible mechanism of DLA was also proposed and observed experimentally [22-26], in which case electrons mostly interact with the transverse field of laser. Even in the case of laser pulse duration of few tens of fs (L≥$\lambda_p$) contribution of DLA along with wakefield (WF) have also been considered [27,28] using pure He target where self-injection (SI) of electrons was applicable. Recently, DLA contribution in laser wakefield accelerator with L>$\lambda_p$ has been established experimentally by Shaw *et al*. [29] using ionization induced injection (III) [30,31] in mixed gas-jet target (He+$N_2$). Further, through detailed PIC simulations it was found that DLA can double the energy of accelerating electrons in a LWFA [32-35]. Importance of DLA in laser plasma acceleration is due to the fact that it can lead to an increase in betatron oscillation of the trapped electrons and thereby very high energy photons through betatron radiation could be generated [36,37]. It may be pointed out here that there are very few experimental reports on pure DLA regime [23,25,26] using long (several hundreds of fs) laser pulses, whereas no reports in the few tens of fs regime, and hence further investigations would be of interest.

In this paper, we present an experimental investigation on electron acceleration using Ti:Sapphire laser pulses of ~55 fs duration (peak power~18 TW, intensity~$5\times10^{18}$ W/cm$^2$) interacting with three different gas targets of He, mixed gas (He+few%$N_2$) and $N_2$. Three distinct regimes of electron acceleration along with the role of DLA have been identified. In case of He, associated with bubble formation and subsequent self-injection of electrons, at a threshold density of ~$5.8\times10^{19}$ cm$^{-3}$ QM electron beam generation was observed, where both wakefield and DLA (Hybrid + self-injection: Regime-1) was found to contribute to the total energy gain of electrons. In case of mixed gas target assisted by ionization induced injection, electron acceleration was observed at a much lower threshold density of ~$2.1\times10^{19}$ cm$^{-3}$. Due to expected weakening of wakefield this could lead to a pure DLA dominated regime (DLA + ionization induced injection: Regime-2). With increase in density to ~$4.1\times10^{19}$cm$^{-3}$, in case of mixed gas target, the role of wakefield increases and the accelerator enters into the hybrid regime, as in case of pure He, but with ionization induced injection (Hybrid + ionization induced injection: Regime-3). 2D PIC simulations using the EPOCH code [38] was performed which also showed clear and distinct features of these three different acceleration regimes, along with generation of surface waves as potential pre-acceleration mechanism for DLA. Further, effect of different acceleration and injection mechanism on the electron beam properties viz. spectrum, beam profile, charge, and pointing stability is also discussed. Such a comparative study in a single experimental set up using three different gas targets (hence applicable different injection mechanisms), thereby identifying the thresholds required for separating the different regimes of acceleration, i.e. pure DLA from hybrid regime, has not been reported earlier, particularly observation of dominated DLA regime with laser pulse duration in the range of several tens of fs.

The remainder of the paper is organized as follows. In Section-2 we provide the details of the experimental set up and other associated parameters. Experimental results on generation of relativistic electron beam and its characteristics using three different gas-jet targets of He, mixed gas (He+$N_2$) and $N_2$ are presented in Section-3. In Section-4, through theoretical analysis and comparison of experimental observations first we identify different acceleration mechanisms i.e. hybrid (wakefield + DLA) and pure DLA applicable at different plasma densities for different gas-jet targets, and is presented in sub-section-A. Subsequently, we corroborate the above analysis with PIC simulations using code EPOCH described in sub-section-B. Next, in Section-5 we discuss the role of different acceleration and injection mechanisms on electron beam properties. Finally, summary and conclusion is presented in Section-6.

## 2. Experimental set up:

A schematic of the experimental set up is shown in Fig. 1. Ti: Sapphire laser pulses (central wavelength of 800 nm) of ~55 fs were focused to a spot of ~25×12.5 μm (radius at $1/e^2$) using f/5 optics along 1.2 mm length of three different gas targets of He, mixed gas (He+few%$N_2$) and $N_2$ (plasma density ~2-7.1×$10^{19}$ $cm^{-3}$) [21,26]. Considering 50% of total energy inside focal spot, the laser pulse provides a total power (P) of ~18 TW and intensity of ~5×$10^{18}$ W/$cm^2$ at focus which corresponds to $a_0$=1.5. The ASE (Amplified Spontaneous Emission) and pre-pulse contrast of the laser pulse was better than $10^{-9}$ (at ~1 ns) and $10^{-7}$ (at 11 ns) respectively. A well characterized gas jet nozzle was used [39,40]. Electron densities in case of He and $N_2$ gas-jets were calculated using the corresponding atomic gas densities for a given pressure and $2^+$ and $5^+$ ionization states respectively. Electron beam spectrum was recorded using a circular magnetic spectrograph (B~0.45 T, diameter~5 cm), with a resolution of ~34% at 30 MeV and ~67% at 60 MeV (for 10 mrad beam). A circular aperture of 10 mm diameter placed close to the entrance of the magnet provided an acceptance angle of ~36 mrad. Phosphor screen (DRZ-high), kept at a distance of ~33 cm from the gas jet target, was used to detect the electron beam and were imaged onto a 14 bit CCD camera. Electron beam profiles were recorded directly on the phosphor screen without magnet in path. The divergence and pointing stability of the electron beams from various gas-jet targets were estimated respectively in terms of angle determined by the ratio of transverse size or position of the beam with respect to the mean on the phosphor screen to the distance of phosphor screen from the source (~33 cm). The resolution of the detection system, limited by imaging set up and the pixel size of the CCD camera, was ~145 μm and corresponds to an angular resolution of <1 mrad at the phosphor screen.

## 3. Experimental results:

The experimental study was performed using fixed laser pulse duration of ~55 fs and intensity of ~5×10$^{18}$ W/cm$^2$ interacting with three different gas targets of He, mixed gas and N$_2$ in the same experimental set up. Similar to our earlier observations [21,26,41], in the present experimental conditions also a regime suitable for highly reproducible and stable generation of electron beams could be identified for all the three gas-jet targets used. Fig. 2 (a) shows a typical dispersed electron beam generated from He gas target at a density of ~5.8×10$^{19}$ cm$^{-3}$ and Fig. 2 (b) shows the corresponding spectra. Corresponding statistics of shots is shown in Fig. S2 (see supplementary). The peak energy (E) is ~28 ± 4 MeV with maximum energy extending upto ~46 MeV (energy corresponding to 10% of the peak flux in the spectrum) with a shot to shot jitter of ± 6 MeV (i.e. ~26 %). In this case, the spectra are mostly quasi-monoenergetic with a mean energy spread ($\Delta E/E$) of ~64% ± 26% containing ~5 pC charge above 7 MeV. The spectra were fitted with a Gaussian function and the FWHM width gives the energy spread of electron beam. Without magnet in path, collimated electron beams were recorded on the phosphor screen having FWHM divergence in the range of ~14-21 mrad (average: ~16 mrad) and ~9-18 mrad (average: ~13 mrad) in the horizontal and vertical directions respectively. A typical electron beam profile, having a FWHM beam size of ~5 mm (~14 mrad) and ~3.3 mm (~10 mrad) respectively in the horizontal and vertical directions, is shown in the inset of Fig. 2 (b). As observed, electron beams generated have an elliptical profile with an average ellipticity of ~1.3. Further the pointing stability of electrons (shown in inset of Fig. 2 b) was found to be ~24×9 mrad in the horizontal and vertical direction respectively.

Next, mixed gas target was used by mixing few percent of $N_2$ with He gas. For the fixed laser pulse duration of ~55 fs electron beam generation was observed at a threshold density of ~$2.1\times10^{19}$ cm$^{-3}$, comparatively lower than He threshold density. In this case also stable generation of electron beams was observed, and typical raw spectra recorded in consecutive shots at different plasma densities are shown in Fig. 2 c. Fig. 2 d shows the corresponding spectra where average of three shots at 2.1, 4.1-4.3$\times10^{19}$ cm$^{-3}$ and average of two shots at other densities are shown. As may be noted, although mostly the spectra is quasi-thermal, QM feature was also observed in once in a while as shown in Fig. 2 c (iii) & (vii) and have not been considered while showing average spectra in Fig. 2 d. The spectra shown in Fig. 2 d are recorded consecutively, and effect of change in plasma density is clearly visible, and in fact at the end of the series similar feature was reproduced when density was brought back to lower density regime of ~$2.1\times10^{19}$ cm$^{-3}$ (Fig. 2 c xiii), which indicates a signature of very good stability of the accelerator. For all the spectra shown in Fig. 2 (c) for mixed gas target total charge is ~20-22 pC above 7 MeV, and with change in density, flux in the different energy regime varied. The maximum energy of the electrons varied from ~55 MeV at lower density of ~$2.1\times10^{19}$ cm$^{-3}$ to a maximum of ~90 MeV at ~$4.2\times10^{19}$ cm$^{-3}$ [Fig. 2c (iv) - (vi)], and decreased to ~45 MeV with further increase in the density [Fig. 2c (ix) - (x)]. Maximum energy observed was similar to that of He in similar density regime however QM feature was absent.

Further, we also performed experiment using pure $N_2$ gas target in the same experimental set up. Similar to the mixed gas target with pure $N_2$ also a threshold density of ~$2\times10^{19}$ cm$^{-3}$ was observed for electron beam generation, and a typical electron spectra recorded are shown in Fig. 2 (e) and Fig. 2 (f), showing quasi-thermal feature. Electron beam charge above 7 MeV was ~22 pC. Statistics of shots showing stable generation of electrons is shown in Fig.S3 (see supplementary). Average maximum energy was ~40 MeV with a shot to shot jitter of ± 10 MeV (i.e. ~50 %). With $N_2$ gas target also collimated electron beams were recorded on the phosphor screen however with a comparatively larger divergence in the horizontal direction i.e. laser polarization direction. FWHM divergence was in the range of ~16-29 mrad (average: ~20 mrad) and 11-20 mrad (average: ~14 mrad) in the horizontal and vertical directions respectively. A typical electron beam profile is shown in the inset of Fig. 2 f, having a FWHM beam size of ~6.5 mm (~19 mrad) and ~4 mm (~11 mrad) respectively in the horizontal and vertical directions. Average ellipticity of the electron beam was slightly higher to ~1.5. However, compared to He, the electron beam pointing stability reduced and was found to be ~3×4 mrad in the horizontal and vertical directions respectively. It may be clarified here that the electron beam profiles recorded were also found to be stable and reproducible, and observed features are based on average of 8-10 consecutive shots.

**4. Role of different acceleration mechanisms: Wakefield and DLA**

**(A) Identification of different regimes of acceleration**

The experimental study was performed using a fixed laser pulse duration of ~55 fs, with plasma density in the range of ~2–7.1×$10^{19}$ cm$^{-3}$ (L/$\lambda_p$~2.2-4.1, $\omega_0/\lambda_p$~1.7-3.16, P/$P_c$~12-43, where $P_c$ is the critical power for self-focusing=17.4×$n_c/n_e$ (GW) [42], $n_c$ is critical density and $n_e$ is electron density). The above parameters conform to the SMLWFA regime [17-21], where strong wakefield leads to self-injection and acceleration of electrons to relativistic energies. At first electron acceleration was studied using pure He gas, where generation of QM electron beams (Fig. 2 a & b) were observed at a threshold density of ~5.8×$10^{19}$ cm$^{-3}$, which is considered as self-injection threshold in the present experimental conditions. This is consistent with several other reports on QM electron beam generation via wakefield acceleration using similar parameters [17-21]. However, as considered in few earlier [27,28] and also recent reports [29], for the condition of L>$\lambda_p$, one would expect an overlap and interaction of laser field with the injected electrons in the bubble leading to gain in energy from DLA also.

We have estimated the maximum energy gain of electrons for DLA regime of acceleration using the formalism derived in our earlier report [24,26], and also described in supplementary, Fig.S1. Results of the theoretical analysis shows that maximum energy gained by electron from DLA is ~25 MeV ($\gamma$=51) in case of He which is less than the observed maximum energy of ~48 MeV. This indicates a significant energy gain of ~21 MeV from wakefield also, and hence suggests a hybrid regime of electron acceleration with almost equal contributions from both wakefield and DLA. We define this hybrid regime of acceleration with self-injection as 'Regime-1'. Also for the conditions of Shaw *et al* [29], where DLA was considered with wakefield in the blowout regime (i.e. hybrid regime), our theoretical formulation predicts comparable energy gain from DLA (66 MeV) and wakefield (66 MeV). Resulting total energy gain of ~132 MeV is almost equal to that observed in the experiment (i.e. >120 MeV).

Next, with a motivation to study the behavior of the accelerator in a regime below the self-injection threshold with the same laser parameters, where reduction in role of wakefield is expected, we used a mixed gas target (~2.5-7.5% $N_2$ in He). This allowed the accelerator to operate at a lower threshold density of ~$2.1\times10^{19}$ cm$^{-3}$ (~7.5% $N_2$) due to ionization induced injection mechanism. By gradually increasing the backing pressure of He, keeping $N_2$ pressure fixed, plasma density could be varied up to ~$7.1\times10^{19}$ cm$^{-3}$ (~2.5% $N_2$), thereby achieving density both below and above the self-injection threshold of ~$5.8\times10^{19}$ cm$^{-3}$ observed with He.

For a plasma density of ~$2.1 \times 10^{19}$ cm$^{-3}$, a maximum energy gain from DLA is estimated to be ~60 MeV, which is comparable to the observed maximum electron energy of ~55 MeV [Fig. 2c (i, ii, xiii), average spectra shown in Fig. 2d magenta color], which suggests dominant DLA regime of acceleration with ionization induced injection (Regime-2). It is consistent with the fact that at much lower density, compared to the self-injection threshold, increase in $\lambda_p$ would lead to deviation from the matching condition for bubble formation [13]. Hence such a scenario would not support bubble formation and strong wakefield generation.

As discussed above also, another possible acceleration mechanism in laser channels for L>$\lambda_p$ could be SMLWFA. However, in several earlier simulations [22] and also in recent reports [26,29,33] it has been found that in similar conditions DLA could be dominant above a threshold value of P/P$_c$ (>6), as complete cavitation at the front of the laser pulse opposes regular wakefield formation in the remaining trailing part of the pulse. In the present case also, high value of P/P$_c$ (~12) supports DLA at a density of ~$2.1 \times 10^{19}$ cm$^{-3}$ due to weakening of wakefield. Similar to other reports [23,25], recently, we also reported experimental observation of DLA using longer laser pulses of 200 fs (P=7.5 TW, P/P$_c$~9-28) in He plasma, at a threshold density of ~$4 \times 10^{19}$ cm$^{-3}$, with observed maximum energy of ~30 MeV [26]. Comparatively, in the present case, use of higher laser power (18 TW) allowed to achieve pure DLA regime at lower density of $2.1 \times 10^{19}$ cm$^{-3}$, with assistance of ionization induced injection, and hence leading to comparatively higher electron energy of ~55 MeV. Here role of ionization induced injection may be emphasized which allowed to achieve acceleration at lower density through DLA using such a short laser pulse which otherwise could not be effective. Recent simulation study has also shown favorable role of ionization induced injection for DLA [35].

Further, with mixed gas target, role of wakefield could be enhanced with increase in plasma density. At a higher density of ~4.1-4.3×10$^{19}$ cm$^{-3}$ [Fig. 2 c (iv)-(vi)], we observed increase in the maximum energy to ~90 MeV, having almost equal contributions from DLA and wakefield. Since this density also is below self-injection threshold, injection takes place by ionization induced injection, and we define this regime as 'Regime-3'. As stated above also, similar scenario of hybrid acceleration with ionization induced injection having equal contributions from DLA and wakefield was observed in a recent experiment reported by Shaw *et al*. [29] and through simulations [32]. Here, we further explore this acceleration regime by enhancing the wakefield strength by increasing the density up to ~7.1×10$^{19}$ cm$^{-3}$ i.e. above self-injection threshold. The contribution of DLA and wakefield was almost found to be equal in the higher density cases also, however maximum energy of electrons reduced as shown in Fig. 2 c (viii-xii).

To further confirm the role of ionization induced injection in DLA, experiment was also performed using N$_2$ gas target. Similar to mixed gas target, in this case also electron acceleration was observed at a threshold density of ~2×10$^{19}$ cm$^{-3}$ (Fig. 2 e & f). Considering the operation at same lowest density of ~2×10$^{19}$ cm$^{-3}$ the electron acceleration and observed maximum energy could be attributed to DLA (Regime-2) as in the case of mixed target discussed above. There are several reports on electron acceleration in N$_2$, both for L<$\lambda_p$ [43,44] and L>$\lambda_p$ [41,45] with a comparatively lower P/P$_c$~2-4 showing wakefield as the dominant acceleration mechanism. However, Adachi *et al* [46] with higher P/P$_c$ ~13.6 considered role of DLA and reported a cascade acceleration of SMLWFA and DLA.

In summary, using high P/P$_c$ and mixed gas target, we could observe electron acceleration through DLA with ionization induced injection (~2.1×10$^{19}$ cm$^{-3}$), which with increase in density transformed to hybrid regime with ionization induced injection (~$4\times10^{19}$ cm$^{-3}$). Hybrid regime of acceleration with self-injection was observed in case of pure He ($\geq$ $5.8\times10^{19}$ cm$^{-3}$). In Fig. 2 g we plot theoretically estimated maximum energies via DLA and wakefield for the range of plasma density used along with the observed experimental values, showing above discussed three distinct regimes of acceleration. Error bars corresponds to the spectrograph resolutions at respective electron energies, except for a data point shown for density of $2\times10^{19}$ cm$^{-3}$. This corresponds to measurements with $N_2$ gas target (Fig. S3) where shot-to-shot jitter associated with the mean maximum energy was ~50%, and higher than the spectrograph resolution of ~35% at that energy. Therefore, error bar here corresponds to observed shot-to-shot jitter. Various experimental observations on electron energies obtained for different gas targets with respective plasma densities along with applicable acceleration and injection mechanism is summarized in Table.1.

**(B) PIC simulation to verify different regimes of acceleration:**

Now, to support the above described three regimes of acceleration we also performed 2D PIC simulations using EPOCH code [38]. The simulation was performed in a moving window frame with box size of 60×80 μm in longitudinal and transverse directions respectively. Laser pulse of 55 fs (L=16.5 μm) duration, wavelength ($\lambda$) 800 nm, and intensity of $5\times10^{18}$ W/cm$^2$, propagating along X with polarization along Y direction, enters the simulation box from left and interacts with the plasma. The plasma length within the simulation box was modeled with an initial linear density ramp of 100 μm followed by 500 μm of uniform density. Total number of sampling electrons was ~$1.5\times10^7$ and a resolution of $\lambda/30$ was used in both directions. In case of He, at a density of $5.8 \times 10^{19}$ cm$^{-3}$, clear bubble formation (Fig. 3 a) was observed after 1040 fs of laser propagation inside plasma, associated with laser pulse modulation leading to pulse compression up to L~8 µm and strong wakefield amplitude >1 (Fig. 3 b) i.e. above wave breaking limit leading to self-injection of electrons. In the case of mixed gas, at a comparatively lower density of $4 \times 10^{19}$ cm$^{-3}$, during initial propagation a channel like structure tending towards bubble formation was seen (similar to that shown in Fig. 3 e and g) with very mild laser pulse modulation. With further propagation laser pulse modulation was observed leading to initiation of bubble formation (Fig. 3 c and 3 d at a time step of 1740 fs). This also leads to strengthening of wakefield (Fig. 3 c & d) however amplitude is still <1. Hence, injection could be only due to ionization induced injection mechanism. At further lower density of $2.1 \times 10^{19}$ cm$^{-3}$ in the case of mixed gas target, only channel formation with betatron oscillation of electron are observed (Fig. 3 e). At this stage, laser pulse modulation is negligible with strength of wakefield <1 is observed (Fig. 3 f). At still lower density of $2 \times 10^{19}$ cm$^{-3}$, in case of N$_2$, an elongated laser channel formation (with almost negligible signature of bubble) was observed after 1040 fs of laser propagation (Fig. 3 g). In this case laser pulse modulation was also absent and hence very weak wakefield amplitude <<1 (Fig. 3 h) was observed, so electron injection could be only due to ionization induced injection. In all the above cases, injected electrons have an overlap with the laser electric field and hence betatron oscillation was clearly observed (Fig. 3 a, c, e, & g), suggesting role of DLA. The gradual decrease in wakefield amplitude with decrease in density leads to transformation from a hybrid regime to DLA dominated regime, as discussed above in the context of experimental observations. It may be noted that, hybrid regime of acceleration observed here with betatron oscillations inside bubble is similar to that also reported by Shaw *et al* (c/f from Fig. 9 in Ref. 29). Betatron oscillation of electrons in case of overlap with laser field has been observed in earlier reported simulation results also [47,48]. Recently, similar hybrid regime of acceleration of electrons for laser pulse duration of few tens of fs regime has also been shown through 2D KLAPS simulations by Feng *et al*. where they also observed clear bubble formation for a laser pulse duration of 30 fs ($L<\lambda_p$), however when the pulse duration is increased to 60 fs ($L\sim\lambda_p$) channel formation is seen with betatron oscillation of electrons (c/f from Fig.1a and b in Ref. 49).

Next, electron energy distribution and relative contribution of DLA and wakefield was estimated from simulation again performed at lower resolution of $\lambda/10$. Plot of normalized longitudinal momentum $P_x$ vs X for He, mixed gas and $N_2$ are shown in Fig. 4 a (i), 4 b (i) and 4 c (i) respectively. In the case of He at a density of $5.8\times10^{19}$ cm$^{-3}$, acceleration of electrons upto ~36-73 MeV, in case of mixed gas target at a density of $4\times10^{19}$ cm$^{-3}$ energy gain of electron upto ~73-110 MeV and in case of $N_2$ at a density of $2\times10^{19}$ cm$^{-3}$, acceleration of electrons upto ~36-55 MeV are observed. The maximum value of energy gain is considered not at the tail of the distribution but at an intermediate point where significant number of electrons are present. Simulation and experimentally observed value of maximum energy gain of electrons in the three cases are consistent. Next, to quantify separate energy contributions of wakefield and DLA, maximum transverse ($\gamma_y$) and longitudinal ($\gamma_x$) energy gain ($\gamma_y = -\int_0^t \frac{2eP_yE_y}{(mc)^2}dt, \gamma_x = -\int_0^t \frac{2eP_xE_x}{(mc)^2}dt$) to the total energy gain $\gamma^2 = 1+\gamma_x+\gamma_y$ were studied for the three cases, as shown in Fig. 4 a (ii), b (ii) & c (ii) respectively. For deducing the contribution of DLA and wakefield the maximum value of $E_x$, $E_y$, $P_x$ and $P_y$ were taken at each time step of 20 fs, in which case all the macro electrons present in the simulation box was considered and no sampling was done to reduce the number of electrons. In case of He, both wakefield and DLA has equal contributions of ~60% and ~40% respectively with self-injection (Regime-1). In case of mixed gas target, at comparatively lower density, wakefield contribution reduces to ~20-25%, in a hybrid regime but with ionization induced injection (Regime-3). In case of $N_2$ at further lower density, wakefield contribution is very negligible, ~5-10% only, and therefore describes a pure DLA in channel with ionization induced injection (Regime-2). It may be noted that as the contribution of DLA increases towards lower density in case of mixed and $N_2$ targets, bunching of electrons is also seen to be prominent in the $P_x$ vs X plot, Fig. 4 b (i) & c (i). Such bunching of electrons is a signature of dominant DLA mechanism as suggested and observed earlier also [22,47,48]. Similar effect was also seen in our earlier study with longer laser pulse of 200 fs duration [26]. This is similar to the micro-bunching of electrons in case of free electron lasers (FEL), where interaction of electrons and electromagnetic wave takes place in the undulator. The above observations suggest onset of wakefield at a threshold density of ~$4\times10^{19}$ cm$^{-3}$, thereby transforming from Regime-2 at lower density to Regime-1&3 at higher density. This is consistent with our experimental observations supported by theoretical estimations (Fig. 2g).

Further, an interesting observation of electron density modulations (surface waves) at the channel boundaries with periodicity of ~1 μm was observed in case of mixed and $N_2$ gas targets at lower densities (Fig. 5 b & c). Observed electron densities approaching >0.1$n_c$ at channel boundaries gives a surface wave wavelength ($\omega_{pe}^2/c\omega_0$) of ~2-3 μm. Generation of surface waves have also been observed in earlier reported PIC simulations mostly using longer laser pulses of few hundreds of fs [40-52], however, here we observe for first time using shorter laser pulse duration of ~55 fs in a gas jet target. Presence of such surface wave in case of mixed and $N_2$ targets supports the role of DLA [52] as it has been suggested that electrons trapped in surface waves are pre-accelerated to relativistic energies and further accelerated by DLA [50,52]. It may be noted that, in case of He (Fig. 5 a), operating at higher densities, very faint surface wave modulations were observed only at the channel boundaries and not at the bubble boundaries. Similar feature is also observed in case of mixed gas target for density of $4\times10^{19}$ cm$^{-3}$ (Fig. 3 c & d). This could be attributed to the strong modulation of the laser pulse in case of higher densities leading to reduction in laser field strength in the back of the laser pulse (range of 275-280 μm of the propagation distance Fig.3a, which is suitable for bubble formation and wakefield acceleration and reduces contribution from DLA.

**5. Discussion on role of acceleration mechanism on electron beam parameters:**

Next, we discuss the effect of three distinct regimes (Regime: 1, 2, 3) on electron beam properties viz. spectrum, charge, beam profile and pointing stability. In case of Regime-1, observed in He target, generation of QM electron beams (Fig. 2 a & b) is typical for bubble regime of acceleration as reported by various groups [17-20] and also observed in our earlier report [21]. Whereas, for Regime-2&3, observed with mixed (Fig. 2 c & d) and $N_2$ (Fig. 2 e & f) gas target, the spectra are quasi-thermal with almost 100% energy spread, suggesting role of ionization induced injection in Regime-3 compared to only self-injection in Regime-1. Another manifestation of ionization induced injection over self-injection is the observed increase in the beam charge above 7 MeV from ~5 pC in case of He to ~22 pC in case of mixed gas and $N_2$ target [30,31]. Typical electron beam profile recorded (inset Fig. 2 b & f), showed ellipticity of 1.3 in case of He (Hybrid: Regime-1), and a comparatively larger divergence and ellipticity of ~1.5 in case of $N_2$ (DLA: Regime-2) with beam profile elongation along laser polarization direction. Further, it was also found that DLA with ionization induced injection in case of $N_2$ (Regime-2), leads to better pointing stability of electron beams (~3×4 mrad) compared to the hybrid with self-injection (Regime-1) in case of He (~24×9 mrad) (Fig. 2 b & f). This could be attributed to the fact that uncontrolled injection in a larger bubble volume in case of self-injection (He) with further interaction with laser field will lead to larger pointing variation, compared to ionization induced injection ($N_2$ target) where injection primarily occurs along the laser axis.

Here, it would also be necessary to consider the different electron beam Twiss parameters i.e. emittance and its effect on propagation and measurement. It has been shown that in laser plasma acceleration interaction with laser field leads to increase in the electron beam emittance [47,53]. Due to significant contribution of DLA in the present experimental conditions, emittance in the laser polarization direction would be larger compared to that in the perpendicular direction, leading to elliptical profile of the electron beams [29,53]. This is also evident in simulation (Fig.5) which shows transverse oscillation and momentum gain of electrons and the effect is more pronounced in case of $N_2$. Mangles *et al* [53] had earlier shown that the ellipticity in the electron beam in laser plasma acceleration is due to enhanced emittance only as the eccentricity was found to increase with the propagation. Measurement of emittance of electron beams and estimation of its effect on evolution and electron beam parameters at different locations in laser plasma acceleration is a complex problem and beyond the scope of the present research work [54].

Other factor to be considered in the ellipticity is laser focal spot itself, as also was the case in the present experiment. However, observed ellipticity in electron beam could not be associated with it as the bubble and channel formation in the plasma also relies on the laser evolution in the initial stage of propagation which could significantly modify the laser focal spot inside plasma due to non-linear processes. Various earlier studies also reported observation of elliptical electron beam profile and showed that it had no correlation with the elliptical laser focal spot [53,55]. Further, interaction and propagation of laser beam in different gas-jet targets and factors affecting it could also be of significance. Here it would be important to discuss effect of laser ASE contrast ratio. The laser intensity of $\sim 5\times10^{18}$ W/cm$^2$ used in the present experiment was much higher than that required for complete ionization of He (He$^{2+}$ threshold is $\sim 8.8\times10^{15}$ W/cm$^2$) and removal of N$_2$ outer five electrons (N$_2^{5+}$ threshold is $\sim 1.5\times10^{16}$ W/cm$^2$). Consequently, plasma formation with corresponding density would take place at the foot of the laser pulse with which main peak of the laser pulse interacts. However, there is always possibility of gas-ionization and pre-plasma formation by the pedestal i.e. ASE present in the laser pulse. Role of pre-plasma in laser plasma acceleration has been an important factor [56,57]. Hydrodynamic expansion of the pre-plasma would lead to decrease in the final plasma density to the moment when the maximum laser intensity is reached, and also affects laser propagation [56]. We have also found role of pre-plasma in our earlier studies and observed generation of high-quality, collimated and stable electron beams when no pre-plasma was present verified using shadowgraphy [21,41]. In the present case also for the ASE contrast of $<10^{-9}$, the corresponding associated intensity was not sufficient for pre-plasma formation. Mangles *et al.* [57] has also studied role of pre-plasma in similar experimental conditions and found that formation of pre-plasma due to comparatively poor ASE contrast leads to generation of multiple electron beams with large pointing stability. Further, another important factor which could affect the laser interaction with plasma is the laser filamentation and ionization induced defocusing (in case of N$_2$) and hence modification of the peak laser intensity estimated using vacuum focal spot.

In our experimental conditions we have observed self-focusing and guiding of laser both in case of He and $N_2$ targets and no detrimental effect of ionization induced defocusing has been found in case of $N_2$ thanks to applicable non-linear focusing processes [21,41]. Relativistic self-focusing leads to higher laser intensity achieved inside plasma i.e. higher than the vacuum intensity. Any filamentation of laser would lead to formation of multiple electron beamlets [57], which is not the case in the present experiment. Formation of single collimated, directional electron beams corroborate stable propagation of laser inside plasma. Also, our experimental conditions of laser focal spot, high value of $P/P_c$ (~12-43) and the density regime were found to lie in the stable propagation regime as described by Borisov *et al* [58]. In many previous studies also stable focusing and propagation of laser has been observed in high-Z gases [43,44,59].

## 6. Conclusions:

In conclusion, an experimental investigation on electron acceleration using ~55fs laser pulses interacting with three different gas targets of He, mixed gas (He+few%$N_2$) and $N_2$, in the density range of $2-7.1\times10^{19}cm^{-3}$, in a single experimental set up, is presented. Role of DLA and wakefield mechanisms are investigated and hence three different regimes of acceleration are identified: hybrid+self-injection (Regime-1) in case of He, DLA+ionization induced injection (Regime-2) in case of mixed and $N_2$ at lower density, and transforming to hybrid+ionization induced injection (Regime-3) at comparatively higher density. Applicability of different acceleration regimes was supported by 2D PIC simulations which also showed surface wave generation in case of mixed and $N_2$ gas targets, which could be a pre-acceleration mechanism to DLA. Observation of pure DLA with ionization induced injection using short laser pulse of few tens of fs duration in mixed and $N_2$ gas targets have not been reported till date. Role of higher laser power along with high value of $P/P_c$ is suggested as the main factor to establish such a regime.

[17] B. Hidding, K.-U. Amthor, B. Liesfeld, H. Schwoerer, S. Karsch, M. Geissler, L. Veisz, K. Schmid, J. G. Gallacher, S. P. Jamison, D. Jaroszynski, G. Pretzler, and R. Sauerbrey, Phys. Rev. Lett. **96**, 105004 (2006).

[18] E. Miura, K. Koyama, S. Kato, N. Saito, M. Adachi, Y. Kawada, T. Nakamura, and M. Tanimoto, Appl. Phys. Lett. **86**, 251501 (2005).

[19] T. Hosokai, K. Kinoshita, T. Ohkubo, A. Maekawa, M. Uesaka, A. Zhidkov, A, Yamazaki, H. Kotaki, M. Kando, K. Nakajima, S. V. Bulanov, P. Tomassini, A. Giulietti, and D. Giulietti, Phys. Rev. E **73**, 036407 (2006).

[20] C.-T. Hsieh, C.-M. Huang, C.-L. Chang, Y.-C. Ho, Y.-S. Chen, J.-Y. Lin, J. Wang, and S.-Y. Chen, Phys. Rev. Lett. **96**, 095001 (2006).

[21] B. S. Rao, A. Moorti, R. Rathore, J. A. Chakera, P. A. Naik, and P. D. Gupta, Phys. Rev. ST Accel. Beams **17**, 011301 (2014).

[22] A. Pukhov, Z.-M. Sheng, and J. Meyer-ter-Vehn, Phys. Plasmas **6**, 2847 (1999).

[23] C. Gahn, G. D. Tsakiris, A. Pukhov, J. Meyer-ter-Vehn, G. Pretzler, P. Thirolf, D. Habs, and K. J. Witte, Phys. Rev. Lett. **83**, 4772 (1999).

[24] G. D. Tsakiris, C. Gahn, and V. K. Tripathi, Phys. Plasmas **7**, 3017 (2000).

[25] S.P.D. Mangles, B. R. Walton, M. Tzoufras, Z. Najmudin, R. J. Clarke, A. E. Dangor, R. G. Evans, S. Fritzler, A. Gopal, C. Hernandez-Gomez, W. B. Mori, W. Rozmus, M. Tatarakis, A. G. R. Thomas, F. S. Tsung, M. S. Wei, and K. Krushelnick, Phys. Rev. Lett. **94**, 245001 (2005).

[26] D. Hazra, A. Moorti, B. S. Rao, A. Upadhyay, J. A. Chakera and P. A. Naik, Plasma Phys. Control. Fusion, **60,** 085015 (2018).

[27] S. Masuda and E. Miura, Phys. Plasmas **16**, 093105 (2009).
23

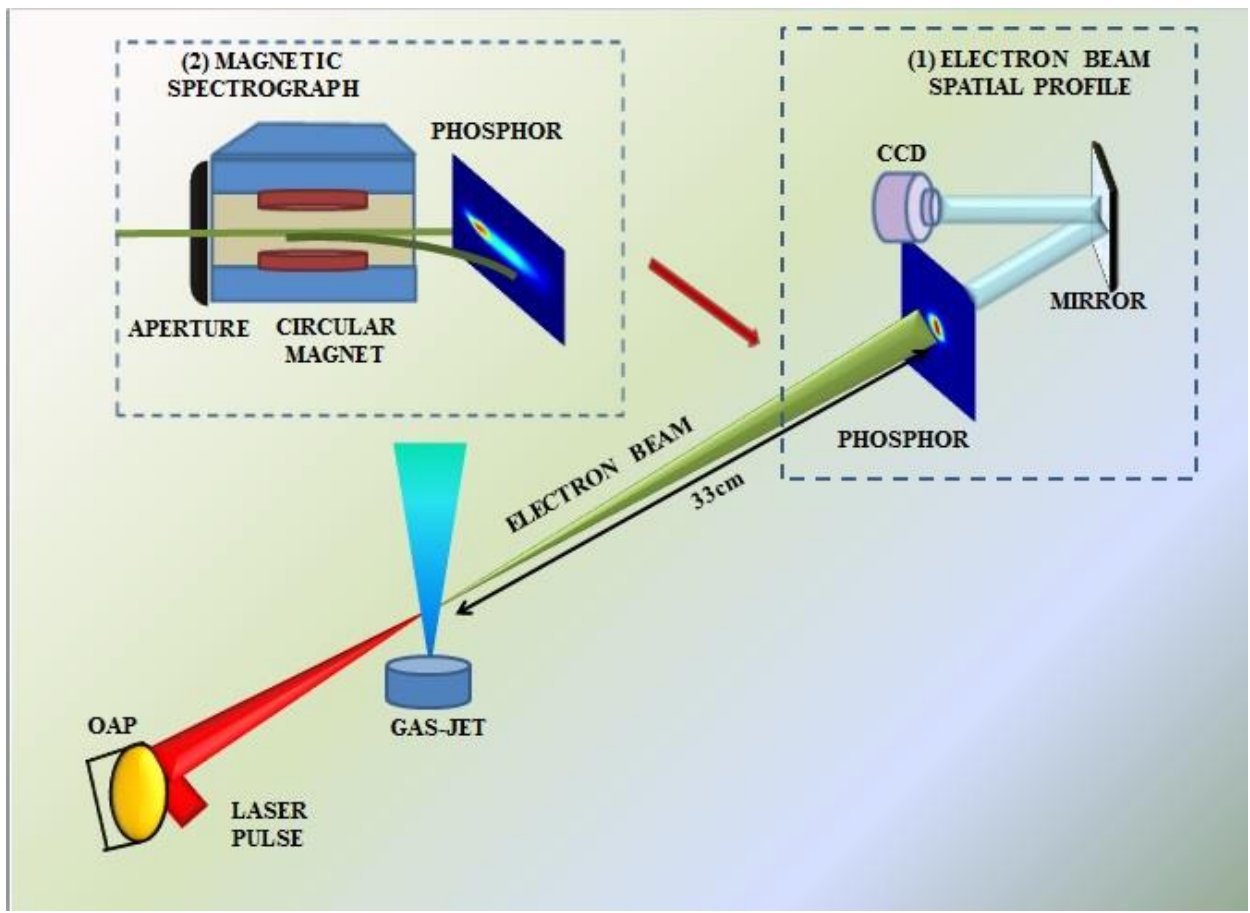

FIG.1 (Color online): Schematic of experimental set up. (a) Recording of electron beam profile using phosphor and CCD camera (b) Recording of electron beam spectra using a magnetic spectrograph inserted in the electron beam path.

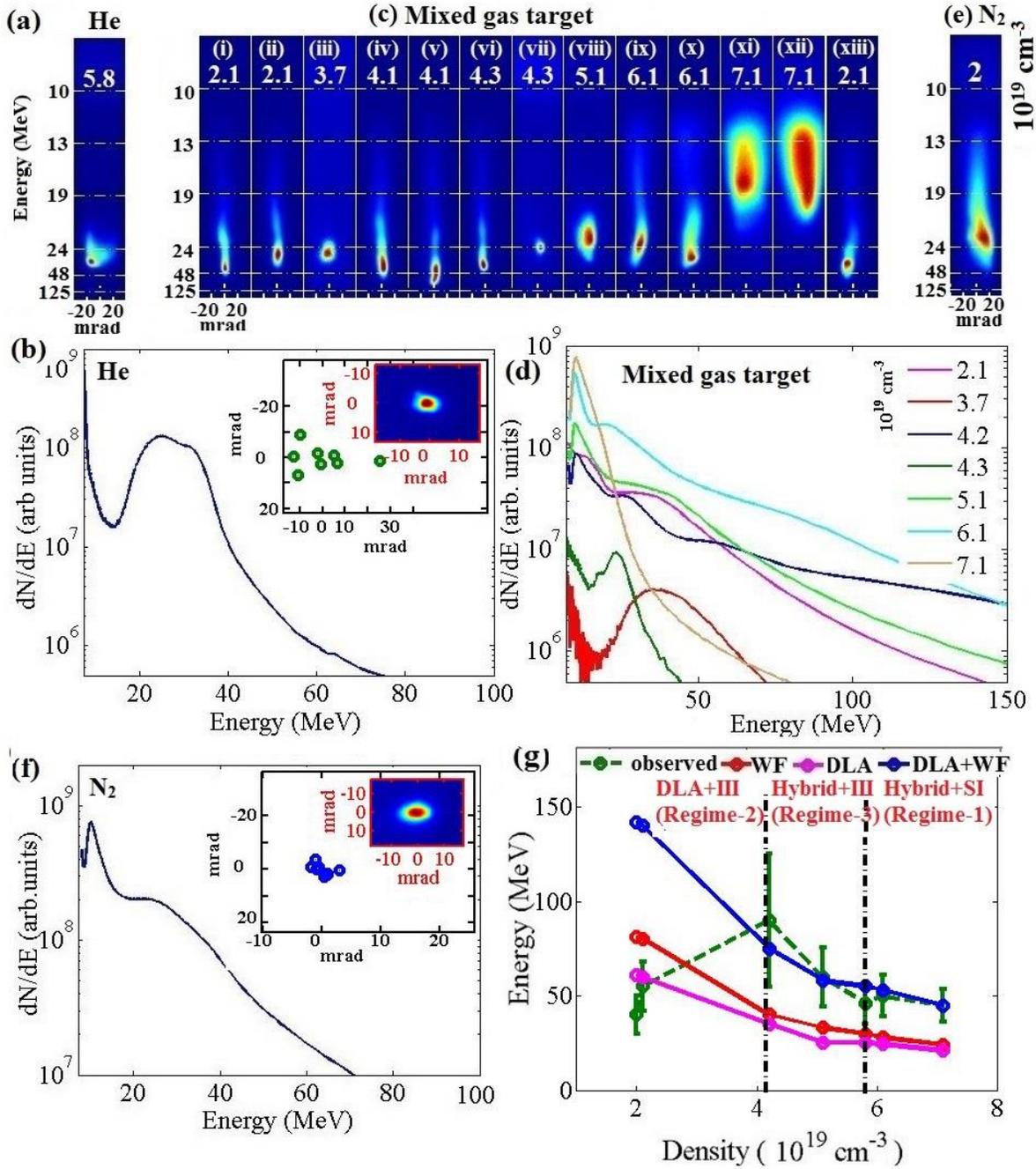

FIG.2 (Color online): (a) Raw images of typical dispersed electron beams and corresponding spectra (a)-(b) for He at density of $\sim 5.8 \times 10^{19} cm^{-3}$; (c)-(d) for mixed (He+$N_2$) gas target at various densities $\sim 2.1 \times 10^{19}$ $cm^{-3}$ ($\sim 7.5\%$ $N_2$) to $\sim 7.1$ ($\sim 2.5\%$ $N_2$) $\times 10^{19}$ $cm^{-3}$; (e)-(f) for $N_2$ at density of $\sim 2 \times 10^{19} cm^{-3}$. Insets in (b) and (f) show pointing stability and typical electron beam profiles (white curves show lineouts). (g) Identification of different acceleration regime by theoretically estimating energy contributions from DLA (magenta) and wakefield (red) at various densities. Experimentally observed maximum electron energies are also shown (green) and compared to DLA + wakefield (blue). Error bars corresponds to spectrograph resolution/shot-to-shot jitter (see also text).

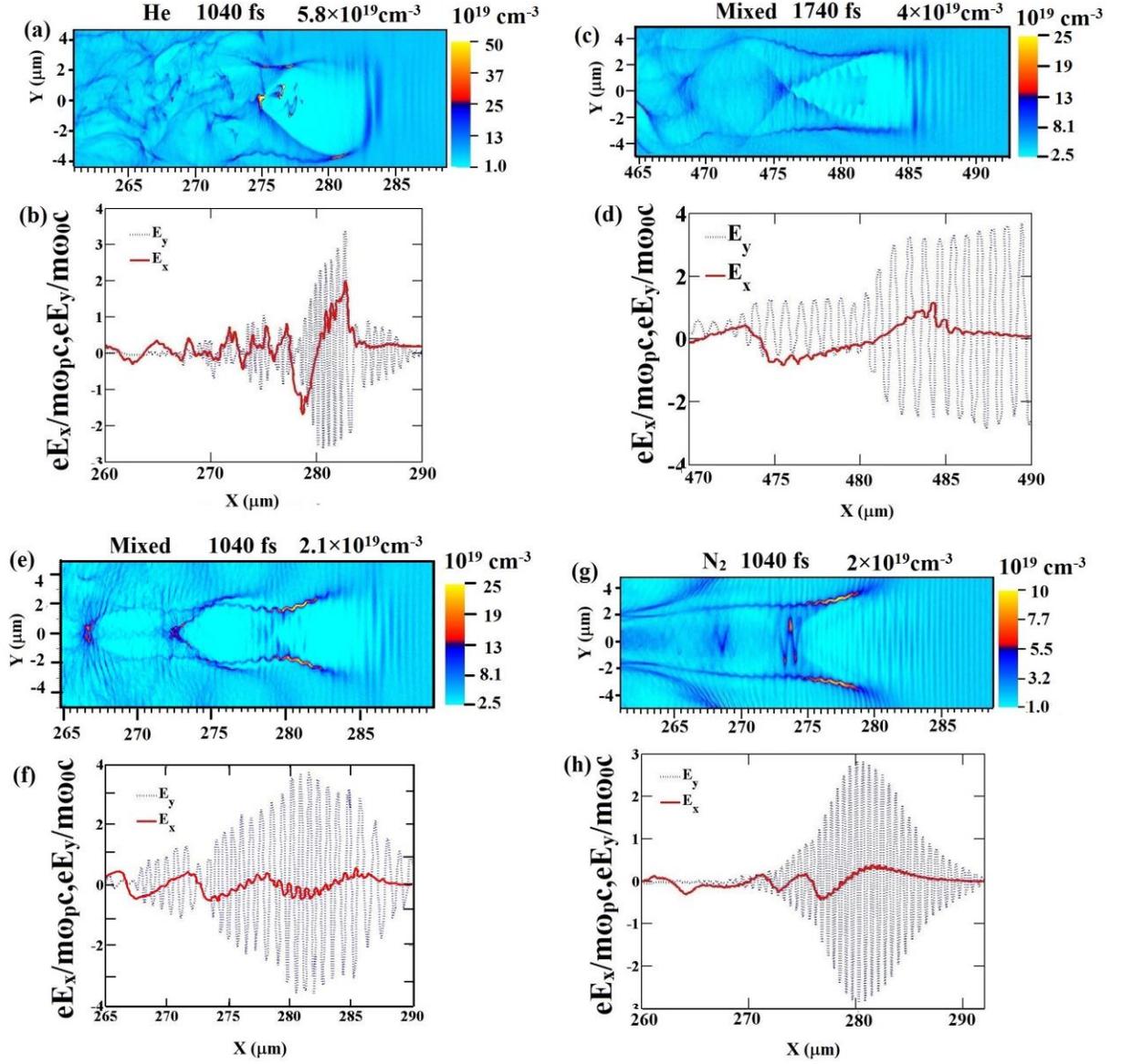

FIG.3. (Color online). Simulation results: Electron density profiles and corresponding lineouts of normalized laser field $E_y$ (blue dot) and wakefield $E_x$ (red solid). (a)-(b) for He at 1040fs for a density of $5.8\times10^{19}$ cm$^{-3}$, (c)-(d) for mixed (He+N$_2$) target at 1740fs for a density of $4\times10^{19}$ cm$^{-3}$, and (e)-(f) for mixed (He+N$_2$) target at 1040fs for a density of $2.1\times10^{19}$ cm$^{-3}$ and (g)-(h) for N$_2$ at 1040fs for a density of $2\times10^{19}$ cm$^{-3}$.

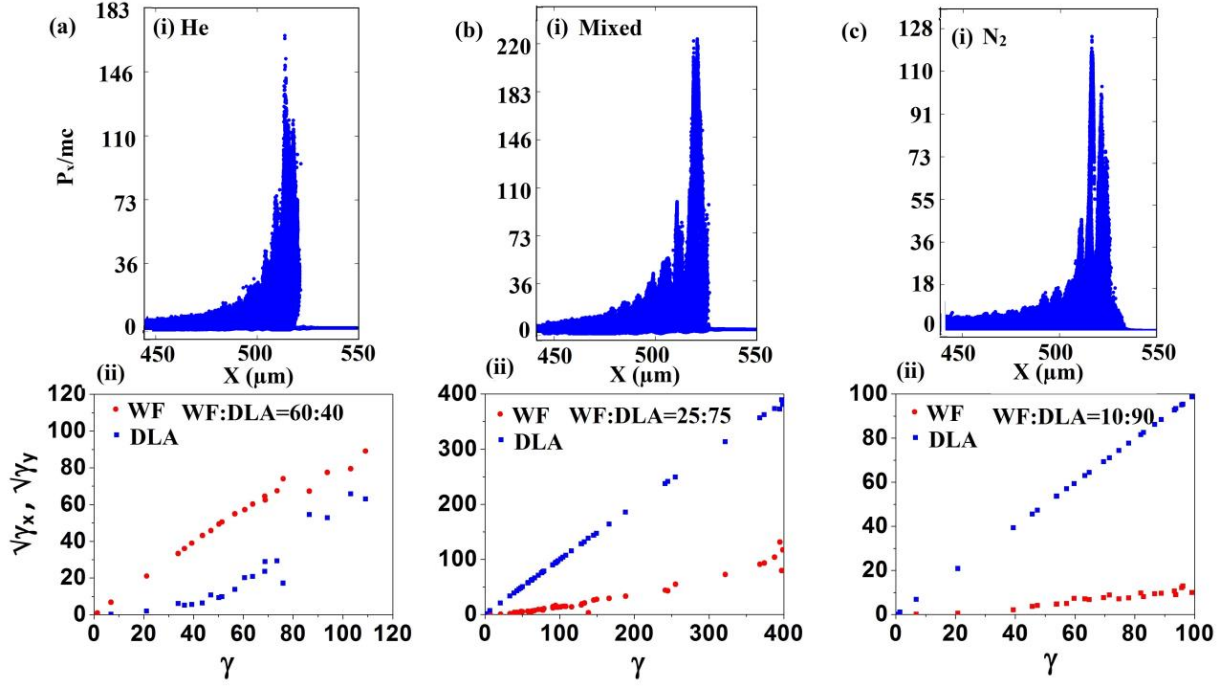

FIG.4. (Color online): Plot of normalized (i) $P_x$ vs X after the propagation of 500μm and (ii) estimation of energy contributions from DLA and wakefield derived from simulations for (a) He at density of $5.8\times10^{19}$cm$^{-3}$, (b) mixed (He+N$_2$) at a density of $4\times10^{19}$cm$^{-3}$ and (c) N$_2$ at a density of $2\times10^{19}$cm$^{-3}$.

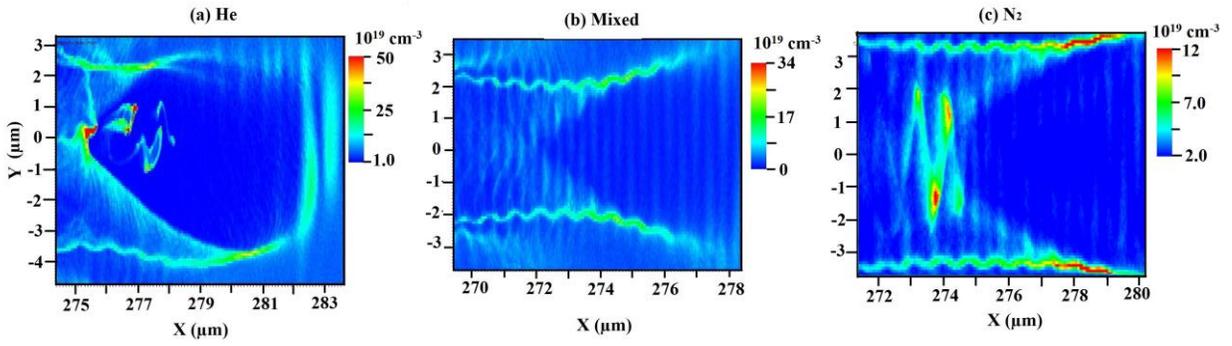

FIG.5. (colour online): Expanded view of electron density profiles simulated at 1040fs (a) He, (b) mixed (He+$N_2$) and (c) $N_2$ targets, showing electron density modulation (SW generation) at channel boundaries in case of mixed and $N_2$.

Table.1: A summary of the experimental observations on electron energies obtained for different gas targets with respective plasma densities along with applicable acceleration and injection mechanism.

| Gas Target | Plasma Density | Maximum Energy (MeV) | Acceleration mechanism | Injection Mechanism |
|---|---|---|---|---|
| **He** | $5.8 \times 10^{19} cm^{-3}$ | 46±10 | Hybrid | Self-Injection |
| **Mixed (He+N$_2$)** | $2.1 \times 10^{19} cm^{-3}$ | 55±13 | DLA | Ionization Induced Injection |
| **Mixed (He+N$_2$)** | $4.2 \times 10^{19} cm^{-3}$ | 90±35 | Hybrid | Ionization Induced Injection |
| **Mixed (He+N$_2$)** | $5.1 \times 10^{19} cm^{-3}$ | 60±15 | Hybrid | Ionization Induced Injection |
| **Mixed (He+N$_2$)** | $6.1 \times 10^{19} cm^{-3}$ | 50±10 | Hybrid | Self-Injection assisted with Ionization Induced Injection |
| **Mixed (He+N$_2$)** | $7.1 \times 10^{19} cm^{-3}$ | 45±9 | Hybrid | Self-Injection assisted with Ionization Induced Injection |
| **N$_2$** | $2 \times 10^{19} cm^{-3}$ | 40±10 | DLA | Ionization Induced Injection |

# Supplementary

## 1. Theoretical electron energy gain estimation from DLA:

Detailed theoretical analysis of betatron resonance acceleration, i.e. direct laser acceleration (DLA), of electrons was reported by Tsakiris *et al*. [1] using plane uniform laser field with linear polarization along x-direction and propagating along z-direction, where an estimation of the maximum energy gain of trapped electrons in the laser field have been derived. In case of betatron resonance acceleration energy of the accelerated electrons $\gamma$ with phase $\phi$ is given by:

$$\frac{d\gamma}{d\phi} = -\frac{eA_0 v_{xA} \cos\phi}{2mc^2 \left(\omega - \frac{\omega_{b0}}{\sqrt{\gamma}} - kv_z\right)} \quad (1)$$

Integrating Eq. (1) we get $F(\gamma) = -P\sin\phi + C_1$, where F($\gamma$) is given by,

$$F(\gamma) = \gamma - \eta\sqrt{(1-\alpha_0)\gamma^2 - 1} + \eta \cos^{-1}\frac{1}{\gamma\sqrt{1-\alpha_0}} - 2\gamma^{1/2}\frac{\omega_{b0}}{\omega} \quad (2)$$

Here $A_0$ is the electric field amplitude, $v_{xA}$ is the on axis velocity of the electrons, $m$ is the mass of electron, $\omega$ is the laser frequency, $\omega_{b0}$ represents the bounce frequency of the oscillation, $k = (\omega/c)\eta$ is the wave number, $\eta = (1 - \omega_p^2/\omega^2(1+a_0^2/2)^{1/2})^{1/2}$ is the ratio of group velocity of the laser in plasma to that in vacuum, $v_z = c(1 - 1/\gamma^2 - v_{xA}^2/2c^2)^{1/2}$ is the axial velocity of electron, $P = a_0 v_{xA}/2c$, $C_1 = F(\gamma_0) + P\sin\phi_0$ is the integration constant, $\gamma_0$ is the initial energy of electrons, and $\phi_0$ is the initial phase of the wave seen by electron, and $\alpha_0 = v_{xA}^2/2c^2$. The plot of F(γ) vs γ (FIG.S1) shows that the curve at first decreases attains a minimum $F_{min}$ at $\gamma_{opt}$ and then increases. The separatrix (phase space behavior: γ vs phase ϕ) is given by equation:

$$F(\gamma) - F_{min} = P(1 - sin\phi) \qquad (3)$$

The largest value of the right hand side of the above equation is equal to 2P. Therefore, a horizontal line was drawn in FIG.S1 at a height of 2P from $F_{min}$, which cuts the F($\gamma$) curve at two points, corresponding to the maximum and minimum energy acquired by a trapped electron. The plot of F(γ) vs γ for three different gas targets of He (at $5.8 \times 10^{19} cm^{-3}$), mixed (at $4 \times 10^{19} cm^{-3}$) and $N_2$ (at $2 \times 10^{19} cm^{-3}$) are shown in FIG.S1. Results of the theoretical analysis shows that maximum energy gained by electron from DLA is ~25MeV (γ=51) in case of He at density of $5.8 \times 10^{19} cm^{-3}$, ~37MeV (γ=75) in case of mixed gas target at density of $4 \times 10^{19} cm^{-3}$ and ~61MeV (γ=123) for $N_2$ at density of $2 \times 10^{19} cm^{-3}$.

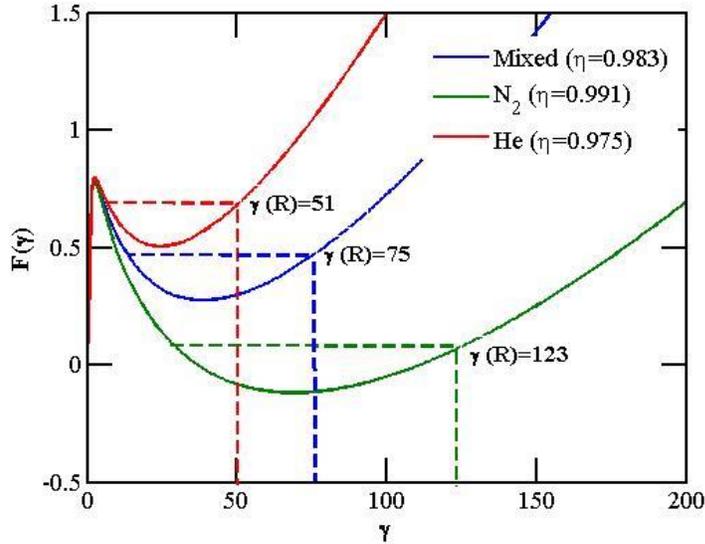

FIG.S1: Variation of F(γ) as a function of γ for $\omega_{b0}/\omega$=0.2, $\alpha_0$=0.03 and η=0.975 (for He) for density $5.8 \times 10^{19} cm^{-3}$, η=0.983 (for mixed) for density $4 \times 10^{19} cm^{-3}$ and η=0.991 (for $N_2$) for density $2 \times 10^{19} cm^{-3}$. γ (R) shows the maximum energy of the electrons.

## 2. Stability of electron beam generation

As observed in our earlier reports [2-4], in the present experiment also, a regime suitable for reproducible and stable generation of electron beams could be identified for all the three gas-jet targets used. In the present experiment, using He gas target, highly reproducible generation of electron beams peak energy (E) of ~28±4MeV and maximum energy extending upto ~46±6MeV were observed at a threshold density of $5.8\times10^{19}$cm$^{-3}$ as shown in Fig.S2. In this case, the spectra are mostly quasi-monoenergetic with a mean energy spread ($\Delta E/E$) of ~64%±26%.

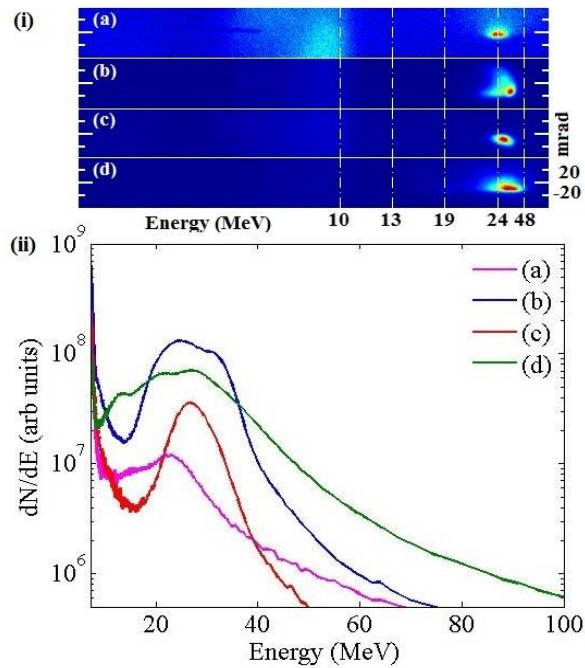

FIG.S2. (i) Raw images of series of dispersed electron beams from He at a density of $5.8\times10^{19}$ cm$^{-3}$. (ii) Spectra of the corresponding raw images.

A series of electron spectra recorded in the present experiment with $N_2$ gas target at a density of $2\times10^{19}$cm$^{-3}$ is shown in Fig.S3. Generation of stable quasi-thermal electron beams with average maximum energy extending upto ~40±10MeV (at 10% of peak electron flux) were also observed in this case.

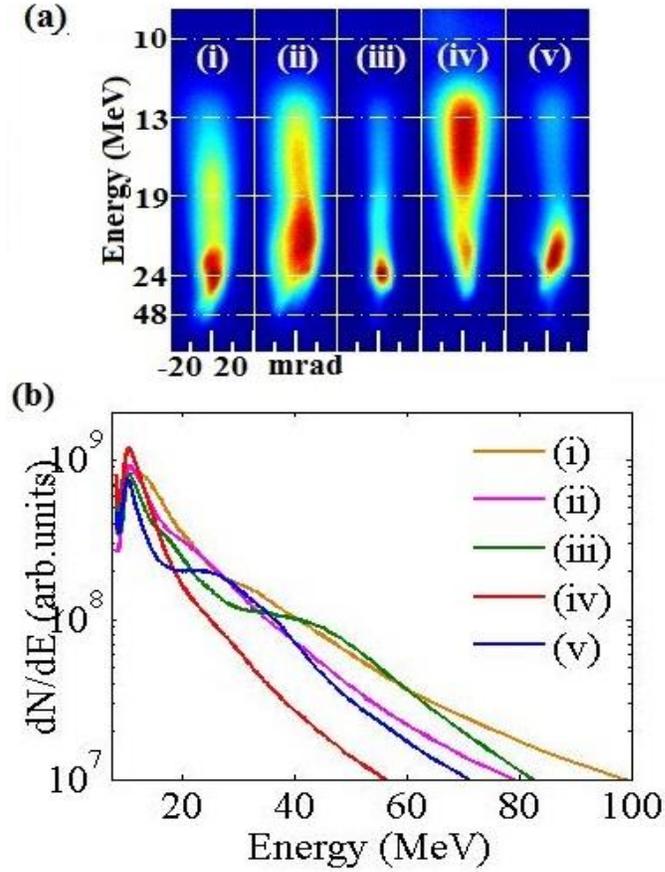

FIG.S3. (a) Series of raw images of typical electron beams from $N_2$ at a density of $2\times10^{19}$ $cm^{-3}$. (b) Corresponding spectra of the electron beams.